\documentclass[12pt,preprint]{aastex}
\usepackage{rotating}

\newcommand{\thebrick}{G0.253+0.016}
\newcommand{\um}{$\mu$m}
\newcommand{\hh}{H$_{2}$}
\newcommand{\Msun}{M$_{\odot}$}

\newcommand{\kms}{km~s$^{-1}$}
\newcommand{\Spitzer}{{\it Spitzer}}

\newcommand{\cms}{${\rm cm}^{-2}$}
\newcommand{\cmc}{${\rm cm}^{-3}$}
\newcommand{\mach}{$\mathcal{M}_{1{\rm D}}$}
\newcommand{\mJybeam}{mJy beam$^{-1}$}

\slugcomment{}

\shorttitle{Turbulence sets the initial conditions for star formation in high-pressure environments}
\shortauthors{Rathborne et al.}

\begin{document}
\title{Turbulence sets the initial conditions for star formation in high-pressure environments}

\author{J. M. Rathborne}
\affil{CSIRO Astronomy and Space Science,  P.O. Box 76, Epping NSW, 1710, Australia; Jill.Rathborne@csiro.au}
\and
\author{S. N. Longmore}
\affil{Astrophysics Research Institute, Liverpool John Moores University, 146 Brownlow Hill, Liverpool L3 5RF, UK}
\and
\author{J. M. Jackson}
\affil{Institute for Astrophysical Research, Boston University, Boston, MA 02215, USA}
\and
\author{J. M. D. Kruijssen}
\affil{Max-Planck Institut fur Astrophysik, Karl-Schwarzschild-Strasse 1, 85748, Garching, Germany}
\and
\author{J. F. Alves}
\affil{University of Vienna, T\"urkenschanzstrasse 17, 1180 Vienna, Austria}
\and
\author{J. Bally}
\affil{Center for Astrophysics and Space Astronomy, University of Colorado, UCB 389, Boulder, CO 8030}
\and
\author{N. Bastian}
\affil{Astrophysics Research Institute, Liverpool John Moores University, 146 Brownlow Hill, Liverpool L3 5RF, UK}
\and
\author{Y. Contreras}
\affil{CSIRO Astronomy and Space Science,  P.O. Box 76, Epping NSW, 1710, Australia}
\and
\author{J. B. Foster}
\affil{Department of Astronomy, Yale University, P.O. Box 208101 New Haven, CT 06520-8101, USA}
\and
\author{G. Garay}
\affil{Universidad de Chile, Camino El Observatorio1515, Las Condes, Santiago, Chile}
\and
\author{L. Testi}
\affil{European Southern Observatory, Karl-Schwarzschild-Str. 2,  85748 Garching bei Munchen, Germany; INAF-Arcetri, Largo E. Fermi 5, I-50125 Firenze, Italy; Excellence Cluster Universe, Boltzmannstr. 2, D-85748, Garching, Germany}
\and
\author{A. J. Walsh}
\affil{International Centre for Radio Astronomy Research, Curtin University, GPO Box U1987, Perth, Australia}

\begin{abstract}
Despite the simplicity of theoretical models of supersonically turbulent, isothermal media, their predictions 
successfully match the observed gas structure and star formation activity within low-pressure  ($P/k<10^5~{\rm K}~{\rm cm}^{-3}$) 
molecular clouds in the solar neighbourhood.  However, it is unknown if these theories extend to clouds in 
high-pressure ($P/k>10^7~{\rm K}~{\rm cm}^{-3}$) environments, like those in the Galaxy's inner 200\,pc Central Molecular Zone (CMZ) 
and in the early Universe. Here we present ALMA 3\,mm dust continuum emission within a 
cloud, \thebrick, which is immersed in the high-pressure environment of the CMZ.  While the log-normal shape and 
dispersion of its column density PDF is strikingly similar to those of solar neighbourhood clouds, there is one 
important quantitative difference: its mean column density is 1--2 orders of magnitude higher. Both the 
similarity and difference in the PDF compared to those derived from solar neighbourhood clouds match predictions of 
turbulent cloud models given the high-pressure environment of the CMZ. The PDF shows a small 
deviation from log-normal at high column densities confirming the youth of \thebrick. Its lack of star formation is consistent with 
the theoretically predicted, environmentally dependent volume density threshold for star formation which is orders 
of magnitude higher than that derived for solar neighbourhood clouds. Our results provide the first empirical evidence 
that the current theoretical understanding of molecular cloud structure derived from the solar neighbourhood 
also holds in high-pressure environments. We therefore suggest that these theories may be applicable to 
understand star formation in the early Universe.
\end{abstract}


\section{Introduction}
Stars form when small ($\lesssim$ 0.1\,pc), dense ($\gtrsim$10$^{4}$\,\cmc) cores in a 
molecular cloud become self-gravitating and collapse (e.g.\,\citealp{Motte98}). Which gas pockets collapse to form 
stars depends on the cloud's internal kinematics and density structure -- theoretical 
studies predict that gravitational collapse, and eventually star formation, will occur once the gas reaches 
a critical over-density ($\rho_{crit}$) with respect to the mean volume density 
($\rho_{0}$). For gas with a 1-dimensional turbulent Mach number, \mach, defined as the ratio of the gas 
velocity dispersion to the sound speed ($\sigma/c_{s}$), the critical density is 
$\rho_{crit} = A\,\alpha_{\rm vir}\,\rho_{0}\,$\mach$^{2}$, where A $\sim$ 1 and $\alpha_{\rm vir}$ is the
virial parameter \citep{Krumholz05,Padoan11}.

Given a critical density, the theoretically predicted rate of star 
formation is obtained by integrating (above $\rho_{crit}$) the volume density probability distribution function ($\rho$-PDF) -- 
the fraction of mass within a cloud at a given volume density. Theoretical models of supersonically 
turbulent, isothermal media show that the $\rho$-PDF follows a log-normal function, because 
the gas experiences a random number of independent shocks that change the volume density by a 
multiplicative factor \citep{Vazquez-Semadeni94}. These theories predict that the dispersion 
of the normalised log-normal $\rho$-PDF ($\sigma_{\log{\rho}}$) increases with \mach\, 
as $\sigma_{\log{\rho}} =\sqrt{\ln{(1 + 3b^2\mathcal{M_{{\rm 1D}}}^2)}}$, where $b$=0.3--1\, \citep{Federrath10}. 
Enhancements that are dense enough to become self-gravitating undergo runaway collapse, 
causing the high-density tail of the $\rho$-PDF to deviate from log-normal 
and follow a power-law \citep{Padoan99,Ballesteros-Paredes11,Kritsuk11}.  
Observations of solar neighbourhood clouds show that their column density PDFs (N-PDFs) 
are also well-described by a log-normal distribution 
\citep{Lombardi08,Goodman09,Kainulainen09,Kainulainen13,Schneider14}. Despite the relative simplicity of 
these models their predictions successfully match the observed gas structure 
within solar neighbourhood clouds \citep{Padoan13}.

In comparison to the solar neighbourhood, the environment within the inner 200\,pc of our Galaxy 
(the Central Molecular Zone, CMZ) is extreme: the volume density, gas temperature, velocity dispersion, 
interstellar radiation field, pressure, and cosmic ray ionization rate are all significantly higher 
\citep{Walmsley86,Morris96,Ao13}. Thus, the CMZ 
provides an ideal laboratory for testing theoretical predictions for cloud structure in an extreme
environment.

To test such predictions, we observed a cloud in the CMZ, \thebrick, using the new Atacama Large 
Millimeter/submillimeter Array (ALMA). With its low dust temperature ($\sim$ 20\,K), high 
volume density ($>$10$^{4}$\,\cmc), high mass ($\sim$ 2$\times$10$^{5}$\,\Msun), and lack of 
prevalent star-formation \citep{Lis94,Lis98,Lis01,Kauffmann13}, it has exactly the properties expected 
for a high-mass  cluster in an early stage of its evolution  \citep{Longmore12,Longmore-ppvi,Rathborne-brick-malt90}. 
Given its location, its detailed study may reveal the initial conditions for star formation in this 
extreme environment. 

\section{Observations}

We obtained a 3'$\times$1' mosaic of the 3\,mm (90\,GHz) dust continuum and molecular line emission 
across \thebrick\, using 25 antennas as part of ALMA's Early Science Cycle 0.  The correlator was 
configured to use 4 spectral windows in dual polarization mode centred at 87.2, 89.1, 99.1 
and 101.1\,GHz each with 1875\,MHz bandwidth and 488\,kHz (1.4$-$1.7\,\kms) channel spacing.  The 
cloud was imaged on six occasions between 29 July -- 1 August 2012. Each 
dataset was independently calibrated before being merged. All data reduction was performed using the 
CASA and Miriad software packages.

\begin{figure}
\includegraphics[angle=90,width=0.95\textwidth,trim=2mm 2mm 2mm 2mm,clip=true]{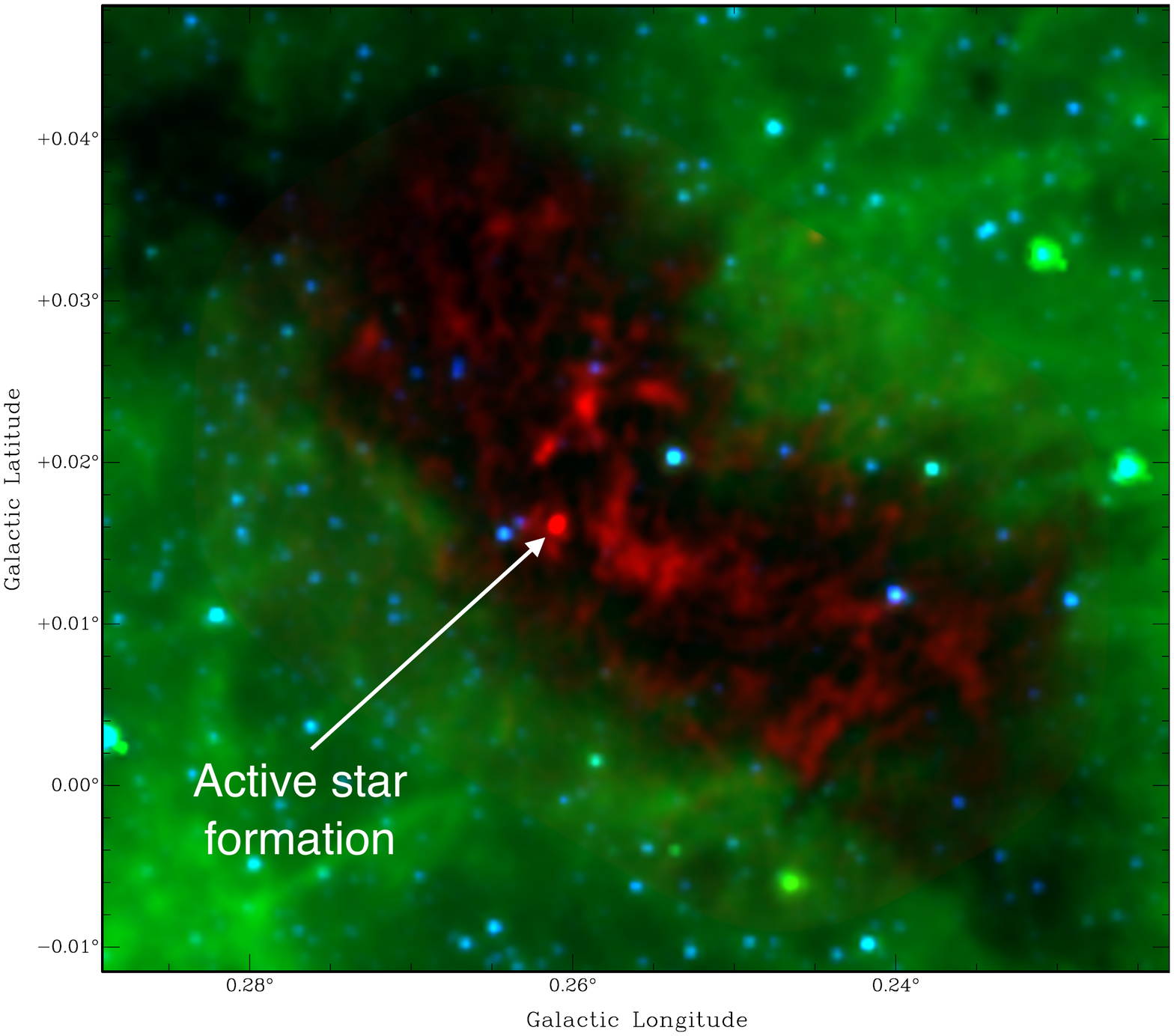}
\caption{\label{image} Three colour image of \thebrick\, (blue is \Spitzer\, 3.6\,\um\, emission tracing stars, 
green is \Spitzer\, 8.0\,\um\, emission tracing the bright Galactic background, while red 
is ALMA 3\,mm emission tracing dust from the cloud's interior; the cloud has an effective radius of 2.9\,pc). The position of a
water maser is marked, which is evidence for active star formation. The cloud is so cold and dense that it is seen as an 
extinction feature against the bright IR emission from the Galaxy. Because ALMA sees through 
to the cloud's interior, we are now able to characterise its internal structure.}
\end{figure}

The continuum image has  a pixel size of 0.35\arcsec,  an angular resolution of 1.7'' (0.07 pc), and a 
 1 $\sigma$ rms  sensitivity of $\sim$ 25$\mu$Jy beam$^{-1}$ (which corresponds to a 5$\sigma$ mass 
sensitivity of $\sim$ 2\,\Msun\, assuming T$_{dust}$=20\,K and $\beta$=1.2).
Because the 90\,GHz spectrum is rich in molecular lines, 
these observations also provided data cubes from 17 different molecular species: combined, they reveal 
the gas kinematics and chemistry within the cloud (see \citealp{Rathborne-overview}).
Figure~\ref{image} combines \Spitzer\, 3.6 and 8\,\um\, images (blue and green, respectively) with the 
new ALMA 3\,mm continuum image (red) of  \thebrick.

\section{Deriving the column density}

{\it {Herschel}} observations show that the dust temperature (T$_{dust}$) within \thebrick\, decreases  
from 27 K on its outer edges to 19 K in its interior \citep{Longmore12}. The ALMA 3\,mm continuum emission 
is enclosed within the region where T$_{dust}$$<$ 22 K,  thus, we assume T$_{dust}$ is 20  $\pm$ 1 K . 

 The ALMA data does not include the large-scale emission ($>$ 1.2') that is filtered out by the 
interferometer. To recover this missing flux, we combined it with a single dish (SD) image that measures the large-scale
emission. Since we do not have a direct measurement of the large-scale 3\,mm emission, we scale 
the 500\,\um\, dust continuum emission ({\it {Herschel}}, 33\arcsec\, angular resolution) 
to what is expected at 3\,mm assuming a greybody where the flux scales like $\nu^{(2+\beta)}$. 
We choose {\it {Herschel}} data as it provides the most reliable recovery of the 
large scale emission: datasets from ground-based telescopes often suffer from large error beams 
or imaging artefacts from data acquisition techniques. While the {\it {Herschel}} data also contains
emission from clouds along the line-of-sight, because \thebrick\, is so cold and dense its emission will dominate.
We choose to scale the 500\,\um\, emission as it is the closest in 
wavelength to 3\,mm, minimising the effect of the assumption of $\beta$. 
Toward the brightest regions in the image, when fitting a greybody to the 3\,mm continuum 
emission derived from each of the 4 individual measurements, we find $\beta$ $\sim$ 1.2--1.5. 
Since the exact value of $\beta$\, across the cloud is unknown, we performed
the image scaling using a range of values for $\beta$\, (1.0, 1.2, 1.5, 1.75, 1.9, 2.0).

\begin{figure}
\includegraphics[width=0.52\textwidth,clip=true,trim=10mm 25mm 5mm 10mm]{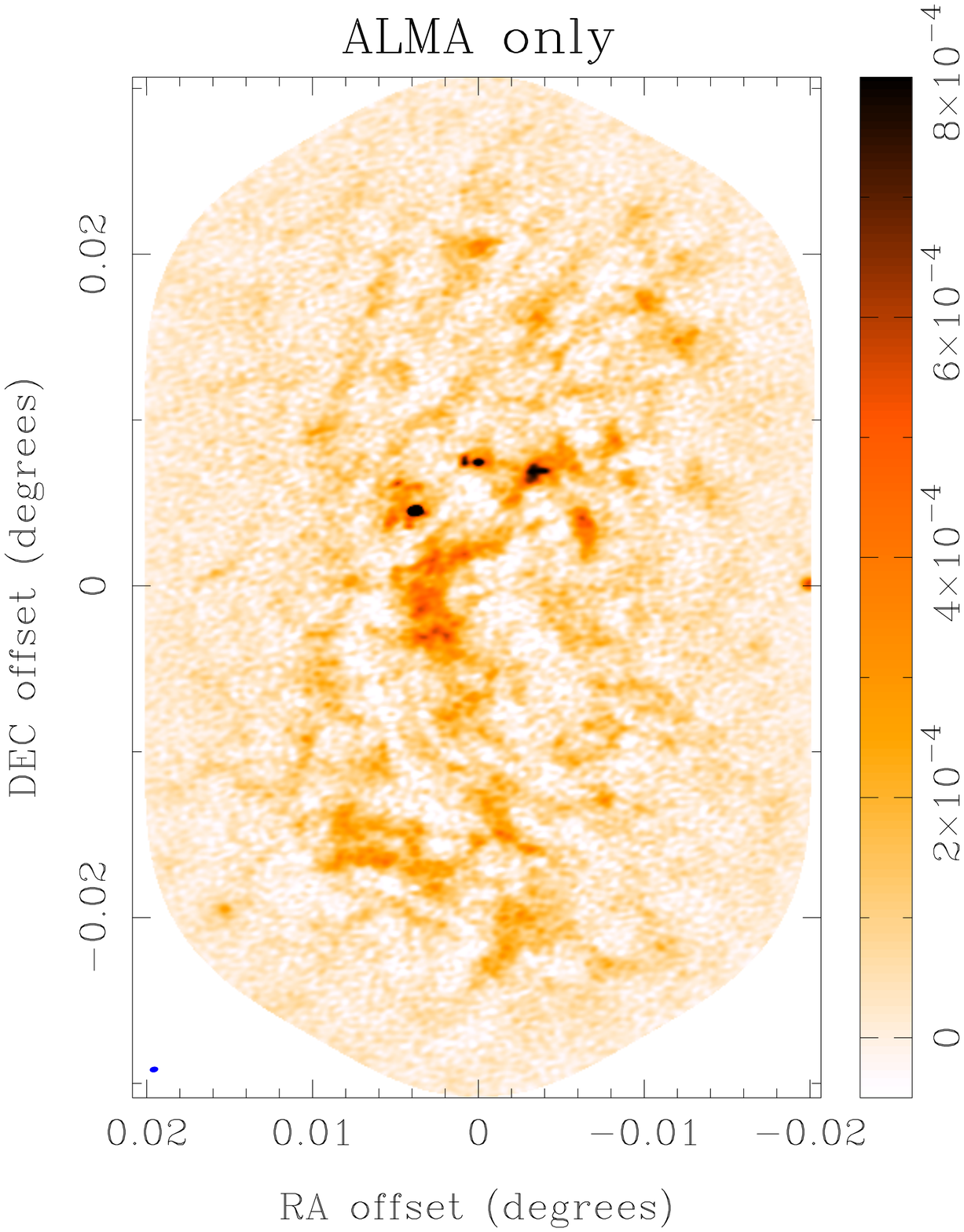}
\includegraphics[width=0.52\textwidth,clip=true,trim=10mm 25mm 5mm 10mm]{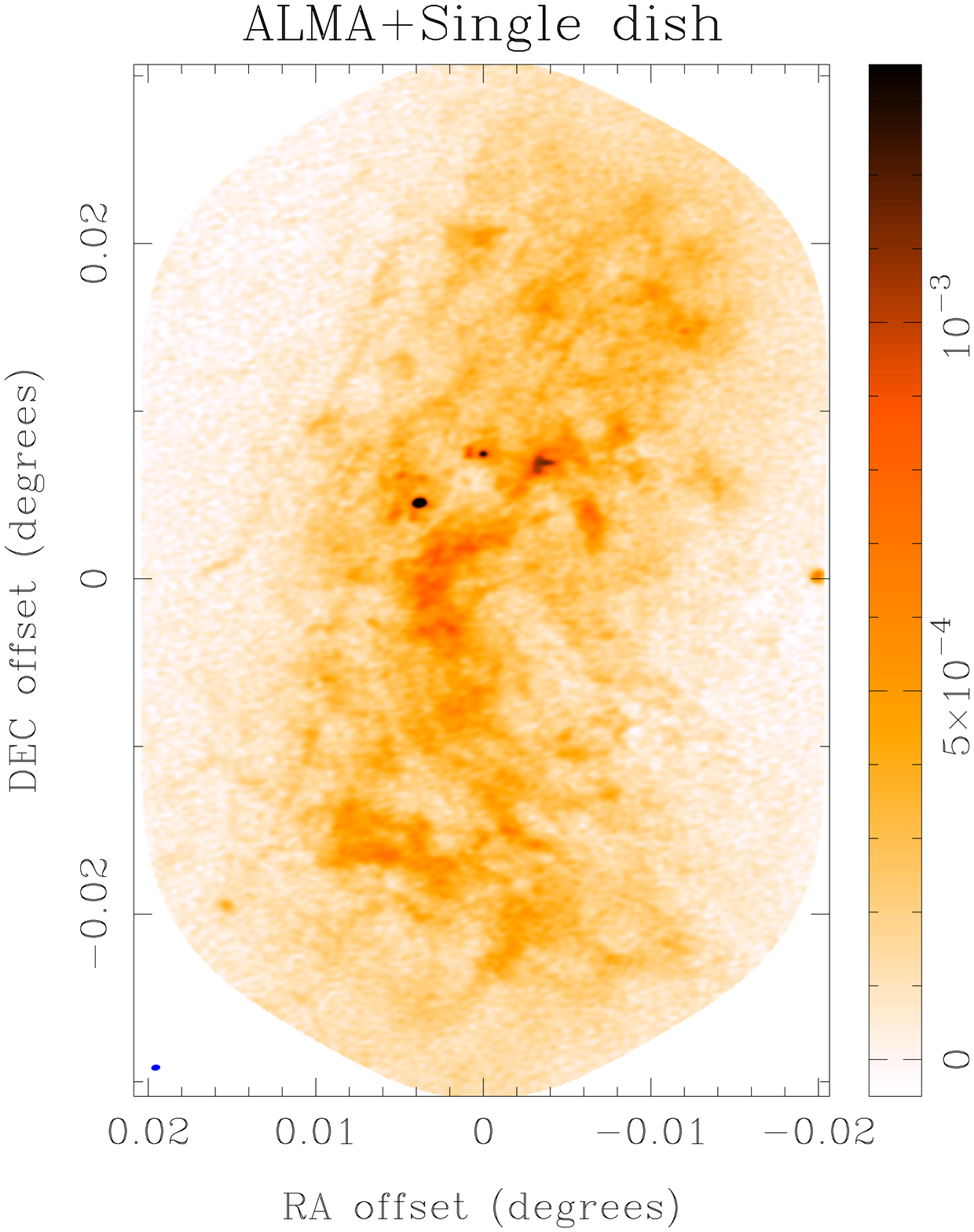}\\
\caption{\label{comparison} Images of the 3\,mm dust continuum emission (in units of \mJybeam) toward \thebrick\, showing the emission
detected in the ALMA-only image (left) and the recovery of the emission on the large spatial scales provided by the inclusion of the 
zero-spacing information (ALMA + single dish, right). These images are shown in equatorial coordinates: the (0,0) offset position in 
R.A. and Dec is 17:46:09.59, $-$28:42:34.2 J2000.}
\end{figure}

The combination of the datasets was performed in CASA via the CLEAN algorithm (see \citealp{Rathborne-overview}). 
Figure~\ref{comparison} shows the ALMA-only continuum 
image (left) and the combined image created using the {\it{Herschel}}\, 500\,\um\, emission assuming a $\beta$ 
of 1.2 (right). The ALMA-only image clearly shows the image artefacts from the missing flux on large scales, 
while the combined image shows the significant improvement and recovery of the large scale emission. The removal of the 
image artefacts justifies the value for $\beta$ $\sim$1.2: higher values underestimate the flux density on large scales 
which does not remove the zero-spacing imaging artefacts. Thus, we assume $\beta$ of 1.2 $\pm$ 0.1. 

Because the 3\,mm emission is optically thin and traces all material along the line of sight, it is proportional to 
the total column density of dust. With T$_{dust}$ of 20 K and assuming a gas to dust mass ratio of 100, dust 
absorption coefficient ($\kappa_{3mm}$) of  0.27  cm$^{2}$ g$^{-1}$ (using $\kappa_{1.2mm}$ = 0.8 cm$^{2}$ g$^{-1}$, 
and $\kappa \propto \nu^{\beta}$; \citealp{Chen08,Ossenkopf94}), and  $\beta$ of 1.2 , the intensity of the 
emission (I$_{3mm}$, in mJy) was converted to column density, N(\hh), by multiplying by  1.9 $\times$ 10$^{23}$ \cms/mJy. 
The uncertainties for T$_{dust}$ and  $\beta$ introduce an uncertainty 
of $\sim$ 10\% for the column density, volume density, mass, and virial ratio. In log-normal fits to the 
N-PDF, there is an uncertainty of $\sim$10\% in the dispersion and  
$\sim$25\% for the peak column density. 

\section{The column density PDF}

\begin{figure}
\centering
\includegraphics[width=0.98\textwidth,trim=3mm 2mm 3mm 0mm,clip=true]{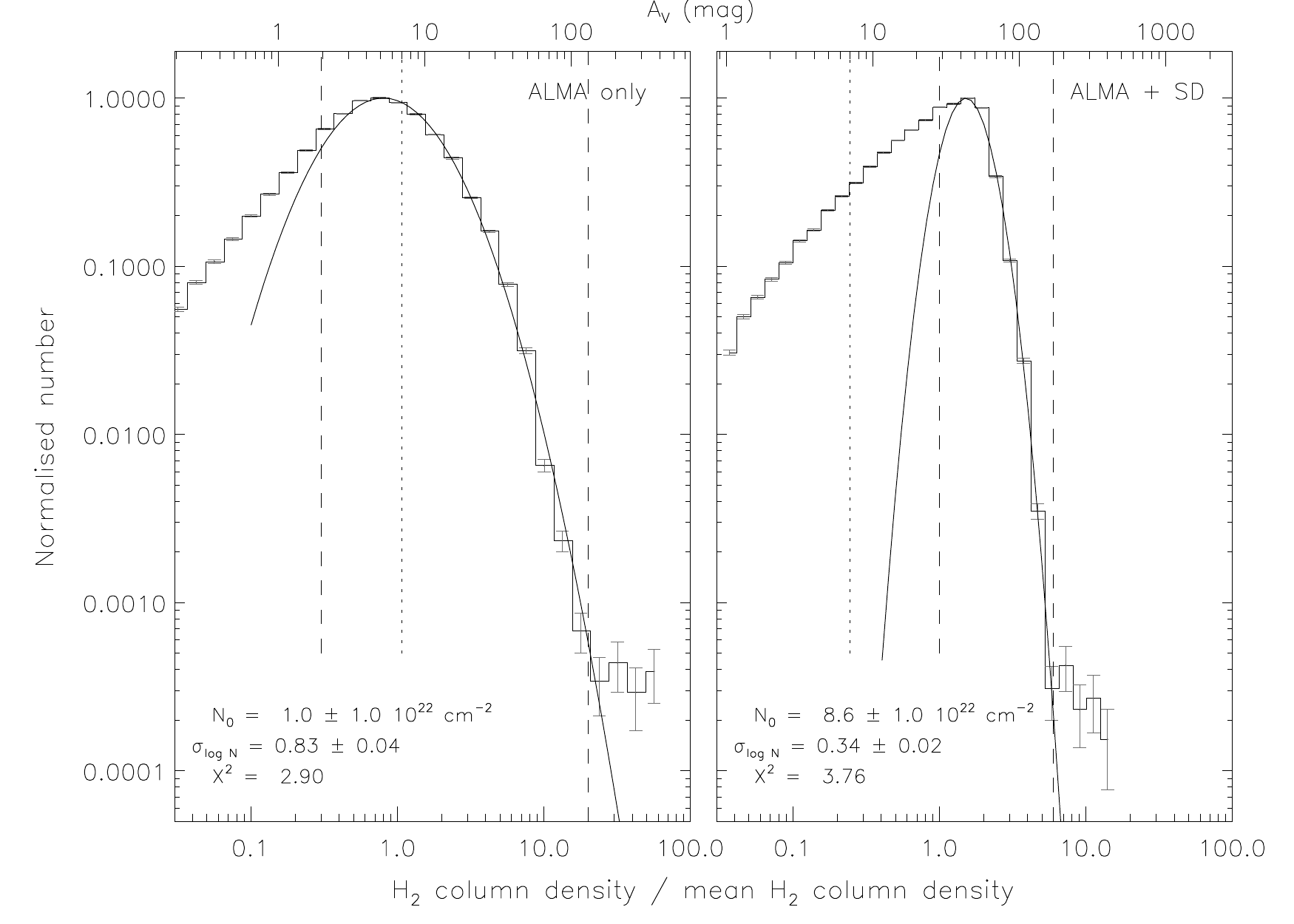}
\caption{\label{pdf} Normalised column density PDFs for \thebrick\, (histograms, {\it{left:}} derived using the 
ALMA-only image,  {\it{right:}} derived using the combined image). The error bars 
show the $\sqrt{number}$ uncertainties. The solid curves are log-normal fits to the PDF: best-fit parameters are labeled. 
Vertical dashed lines show the fit range (the limits mark the approximate point at
which the distributions deviate from log-normal). Vertical dotted lines mark N(\hh)=1.4 $\times$ 10$^{22}$\,\cms. The small deviation at the 
highest column densities traces material that is self-gravitating and corresponds to the only location in the cloud where star formation is occurring.}
\end{figure}

The  sensitivity and angular resolution of the ALMA data allows us to derive the N-PDF for \thebrick\, to 
high accuracy\footnote{Recent work based on 1\,mm SMA observations toward \thebrick\, has also measured its N-PDF,
see \cite{Johnston14}. }. Figure~\ref{pdf} compares the N-PDF derived from the ALMA-only data to that derived 
from the combined image (left and right respectively).  Both N-PDFs are well fitted by a log-normal 
distribution.  When using the combined image, the shape of the N-PDF remains log-normal, however, the derived 
dispersion is narrower and the peak column density higher compared to using the ALMA-only image. These differences 
are expected when including/excluding large scale emission \citep{Schneider14}. 

The deviation from log-normal at low column densities arises from the large-scale diffuse medium 
 and is a common feature in other PDFs (e.g. Lupus I, Pipe, Cor A, see fig.~4 from \citealp{Kainulainen09}).  
Since the ALMA-only image recovers a small fraction of the 
total flux ($\sim$ 18\%), its PDF will characterize the highest contrast peaks within the cloud. Thus, to
make meaningful comparisons between the N-PDF for the \thebrick\, and to  solar neighbourhood 
clouds and theoretical predictions requires the inclusion of the large-scale emission. Thus, we use the values derived from 
that N-PDF (i.e.\,Fig.~\ref{pdf}, right) but report both sets of values for completeness.

There is a small deviation from the log-normal distribution at the highest column densities which indicates 
self-gravitating gas where star formation is commencing.  This high-column density material arises 
from contiguous pixels that exactly coincide  with the location of previously-known water maser emission -- the only evidence 
for star formation within the cloud \citep{Lis94,Kauffmann13}. Because the immediate vicinity of a forming star is heated, this 
deviation may result from a higher temperature in this small region rather than a higher column density.  Nevertheless, in 
either case, this deviation from the log-normal distribution indicates the effect of self-gravity.  

Assuming a dust temperature of 20 K, we calculate the mass of this core (R$\sim$0.04\,pc) to be  72\,\Msun ,  and its 
volume density to be  $>$3.0  $\times$ 10$^{6}$\,\cmc\, (with  T$_{dust}$=50\,K, M=26\,\Msun, and $\rho >$ 1.2 
$\times$ 10$^{6}$\,\cmc ; the density is a lower limit since this core is unresolved).
To assess whether it is gravitationally bound and unstable to collapse or unbound and transient, 
we determine the virial parameter, $\alpha_{\rm vir}$, defined as  $\alpha_{\rm vir}=3k\sigma^2R/GM$,  where $\sigma$ is the 
measured  1-dimensional  velocity dispersion, $R$ the radius, $G$ the gravitational constant, 
$M$ the mass, and  $k$ is a constant that depends on the density distribution \citep{MacLaren88}. For high-mass star-forming 
cores with density profiles $\rho \propto r^{-1.8}$ \citep{Garay07}, $k$ is 1.16.  
 Because the core's associated H$_{2}$CS molecular line emission is unresolved in velocity,  $\sigma$  
$<$1.4\,\kms. Thus, for a mass of  72 \,\Msun\,  $\alpha_{\rm vir}$ $<$  1.1  (for  M=26\,\Msun, $\alpha_{\rm vir}$ $<$ 2.8 ).  
Since  $\alpha_{\rm vir}$ is  $\sim$ 1 , this star-forming core is  likely  gravitationally bound and unstable to collapse.

\section{Discussion}
\subsection{Comparison to solar neighbourhood clouds}

In this section we show that both the similarities and differences in the PDFs for solar neighbourhood and CMZ clouds agree with 
predictions of turbulent models given their environments (for a summary, see Table~\ref{compare} and references therein).

The similarity in the measured dispersions of the N-PDFs ($\sigma_{\log{N}}$=0.28--0.59 in the solar neighbourhood and 
 0.34 $\pm$ 0.03  in \thebrick) is understood by considering their turbulent Mach numbers. The gas temperature in the 
solar neighbourhood and CMZ ($\sim$10\,K and $\sim$65\,K) correspond to  sound speeds  
($c_{s}$) of $\sim$ 0.2 and 0.5\,\kms, respectively. Given the observed velocity dispersions ($\sigma$$\sim$ 2 and $\sim$15\,\kms\, 
respectively),  their \mach\, numbers are $\sim$10 and $\sim$30, which corresponds to predictions of $\sigma_{\log{\rho}} \sim$ 
2.08 and 2.55 for the solar neighbourhood and CMZ respectively (assuming $b$=0.5).  Thus, while the \mach\, for CMZ 
clouds compared to solar neighbourhood clouds is higher, the predicted values for $\sigma_{\log{\rho}}$ differ by only a factor of 
1.2 due to the weak dependence of $\sigma_{\log{\rho}}$ on \mach.

The difference in the mean column densities of the N-PDFs ($N_{0}=0.5$--$3.0\times10^{21}$\,\cms\, in the solar neighbourhood and 
 86 $\pm$ 20  $\times10^{21}$\,\cms\, in \thebrick) is understood by considering the relative gas pressures. The turbulent gas 
pressure is given by $P_{turb} = \rho\, \sigma^{2}$. 
For typical values for solar neighbourhood  ($\rho$$\sim$10$^{2}$\,\cmc; 
$\sigma$$\sim$2\,\kms) and  CMZ clouds ($\rho$ $\sim$10$^{4}$\,\cmc; $\sigma$$\sim$15\,\kms), 
the turbulent gas pressures in units of 
$P/k$ are 10$^{5}$ and 10$^{9}$\,K\,\cmc\,  respectively. The hydrostatic pressure from self gravity ($P_{grav}$) is 
related to the gas surface density ($\Sigma$) through $P_{grav} =  (3/2) \pi G \Sigma^{2}$. Given the surface density of  
 solar neighbourhood  ($\Sigma$$\sim$$10^{2}$\,\Msun\,pc$^{-2}$) and  CMZ clouds 
($\Sigma$$\sim$$5\times10^{3}$\,\Msun\,pc$^{-2}$), 
the respective hydrostatic pressure in units of $P/k$ are also 10$^{5}$ and 10$^{9}$\,K\,\cmc. As $P_{turb} \approx P_{grav}$, the 
 pressures are close to equilibrium on the cloud scale for both environments. 
Because $P_{grav}\sim\Sigma^2$, the condition of hydrostatic equilibrium translates the factor of $10^4$ difference 
in turbulent pressure to a factor of $\sim10^2$ difference in column densities. This explains the factor of 10$^{2}$
difference between the mean column density for solar neighbourhood clouds and the CMZ cloud \thebrick.

The conversion of a N-PDF to a $\rho$-PDF has not been solved conclusively. Theoretical work suggests that the conversion is a  
multiplication by a factor $\xi$, where $\sigma_{\log{\rho}}=\xi\, \sigma_{\log{N}}$\,\citep{Brunt10b}. The uncertainty on $\xi$ is $\sim$15\%\,Ê
for the values of $\sigma_{\log{N}}$\,Êmeasured in solar neighbourhood and CMZ clouds \citep{Brunt10b}\, -- smaller 
than the observed spread of the measured  $\sigma_{\log{N}}$. The relative universality of $\xi$ means that the small relative 
change of the N-PDF dispersions 
($[\sigma_{\log{N}}]_{\rm CMZ}\sim[\sigma_{\log{N}}]_{\rm Solar}$) translates to the same relative change of the 
$\rho$-PDF dispersions, thereby allowing a direct comparison of the measurements to theory. 
Thus, within the uncertainties, the N-PDF of \thebrick\, 
satisfies the theoretical prediction that $[\sigma_{\log{\rho}}]_{\rm CMZ}\sim[\sigma_{\log{\rho}}]_{\rm Solar}$, providing the first reliable
test of turbulence theory in a high-pressure environment. 
Because we neglected magnetic fields, the similarity in the predicted and 
measured dispersions suggests that the thermal-to-magnetic pressure ratios \citep{Molina12} are also comparable in the 
solar neighbourhood and the CMZ (and are likely close to unity). However, future observations of the magnetic fields within the 
high-density material in both environments are needed to confirm this. 

\subsection{An environmentally dependent star formation threshold}

Observations of solar neighbourhood clouds suggest a column density threshold of $\sim$ 1.4 
$\times$ 10$^{22}$\,\cms, above which star formation proceeds with very high efficiency on a 20\,Myr time-scale \citep{Lada10}. 
While the exact interpretation has been questioned \citep{Gutermuth11, Burkert13} subsequent work suggests a `universal' column 
density threshold for star formation \citep{Lada12}.  This empirically-motivated universality is at odds with the volume density 
threshold predicted by theoretical models of turbulence, which depends on $\rho_{0}$ and \mach. 
Despite it accurately describing star formation in 
solar neighbourhood clouds, this `universal' threshold does not hold for the CMZ: the majority of the 
gas has N(\hh) $\gg$ 1.4 $\times$ 10$^{22}$\,\cms, yet it is forming stars 1--2 orders of 
magnitude less efficiently than predicted by this threshold  \citep{Longmore13}. 
The N-PDF for \thebrick\, confirms this result. While the majority of the mass has N(\hh) $>$ 1.4 $\times$ 
10$^{22}$\,\cms, only one region, corresponding to 0.06\% of the total mass, shows evidence for star formation (Fig.~\ref{pdf}). 
This result casts doubt on a universal threshold for star formation of N(\hh) $\sim$ 1.4 $\times$10$^{22}$\,\cms.

This discrepancy can be understood by considering predictions of theoretical 
models of turbulent clouds \citep{Krumholz05, Padoan11}. Using the mean volume density and Mach number 
of the solar neighbourhood ($\rho_{0}$$\sim$10$^{2}$\,\cmc; \mach$\sim$10) and 
CMZ clouds ($\rho_{0}$$\sim$10$^{4}$\,\cmc; \mach$\sim$30), the predicted values for $\rho_{crit}$ are $\sim$10$^{4}$ and 
$\sim$10$^{8}$\,\cmc, respectively.  Empirical studies show that, in solar neighbourhood clouds, column densities of 
1.4 $\times$ 10$^{22}$\,\cms\, correspond to volume densities of $\sim$ 10$^{4}$\,\cmc\, (suggesting a common size scale 
of star forming cores of $\sim$0.2 pc; \citealp{Bergin01,Lada10}). Thus, for the solar neighbourhood, this volume density agrees with 
the prediction for $\rho_{crit}$.

The sole star-forming core in \thebrick, with a derived volume density $> 10^{6}$\,\cmc\, 
conforms to the higher threshold predicted for CMZ clouds.
Its associated molecular line emission (from H$_{2}$CS) traces very dense gas (i.e. $>$ 10$^{7}$\,\cmc), confirming that $\rho \gg 10^{6}$\,\cmc.
While higher resolution observations are required to spatially resolve the core, this derived 
lower limit is consistent with the theoretically predicted, 
environmentally dependent volume density threshold -- orders of magnitude higher than derived for solar neighbourhood clouds.

\subsection{CMZ clouds as local analogues of clouds in the early Universe}

Recent surveys have unveiled rapidly ($100$--$1000~{\rm M}_\odot~{\rm yr}^{-1}$) star-forming 
galaxies \citep{Forsterschreiber09,Swinbank10,Daddi10} at high redshifts ($z>2$), near the peak epoch of 
 star formation \citep{Hopkins06}. Understanding how these  high 
star formation rates can be achieved is one of the main challenges in galaxy formation. Building 
upon both their simplicity and success, models of turbulent cloud structure based on solar neighbourhood clouds have been applied to 
explain the extreme star formation activity in these galaxies \citep{Krumholz12,Renaud12}. However, their low 
turbulent pressures ($P/k<10^5~{\rm K}~{\rm cm}^{-3}$) differ from their high-redshift counterparts by several 
orders of magnitude ($P/k>10^7~{\rm K}~{\rm cm}^{-3}$; \citealp{Swinbank11,Kruijssen13}). 

Given the modest metallicity difference between the CMZ and rapidly star-forming, high-redshift galaxies 
(less than a factor of 2--3; \citealp{Erb06,Longmore13}), CMZ clouds have the potential to be 
used as local analogues of clouds in $z>2$ galaxies \citep{Kruijssen13}. Indeed, CMZ clouds
can be studied in a level of detail that is unachievable for clouds at earlier epochs. 
Our results provide the first empirical evidence that the 
current theoretical understanding of structure derived from solar neighbourhood clouds
also holds in extreme, high-pressure environments. As such, the application of these theories to describe 
star formation in the early Universe may be valid.

\section{Conclusions}

Using the new ALMA telescope to obtain high sensitivity 3\,mm observations, 
we have measured the dust column density with an
extreme cloud in the CMZ, \thebrick, to high accuracy. Our analysis shows that the log-normal shape, dispersion, and mean
of its column density PDF very closely matches the 
predictions of theoretical models of supersonic turbulence in gas of such high density and turbulence.  
The lack of wide-spread star formation throughout the cloud, combined with the fact that 
the PDF shows a small deviation at high-column densities, confirms the youth of \thebrick. 
Our results are consistent with 
the theoretically predicted environmentally-dependent threshold for star formation which  provides a 
natural explanation for the low star formation rate in  the CMZ \citep{Kruijssen14}.  
The confirmation of these models in a high-pressure environment 
suggests that our current theoretical understanding of gas structure derived from solar neighbourhood clouds
may also hold in the early Universe.

\acknowledgements
We thank our ALMA Contact Scientist, Dr Crystal Brogan, for preparing the observations and performing the 
 data calibration. J.M.R acknowledges funding support via CSIRO's Julius Career Award. 
 J.M.R, S.N.L, J.M.D.K, J.B. and N.B. acknowledge the hospitality of the Aspen Center for Physics, 
 which is supported by the National Science Foundation Grant No.~PHY-1066293. J.M.J gratefully acknowledges support from the US 
 National Science Foundation grant AST 1211844. We make use of the following ALMA data: 
 ADS/JAO.ALMA\#2011.0.00217.S. ALMA is a partnership of ESO (representing its member states), NSF (USA) and NINS (Japan), 
 together with NRC (Canada) and NSC and ASIAA (Taiwan), in cooperation with the Republic of Chile. The Joint ALMA Observatory is 
 operated by ESO, AUI/NRAO and NAOJ.

{\it Facilities:} \facility{ALMA, Herschel}


\begin{table}
\caption{\label{compare}Molecular cloud properties within the solar neighbourhood (Solar) and Central Molecular Zone (CMZ). The key properties are marked in bold.}
\vspace{0.2cm}
{\footnotesize{
\begin{tabular}{lllll}
\hline
\multicolumn{2}{l}{}  & Solar & CMZ & References\\
\hline
\multicolumn{2}{l}{Observed:} && & \\
& Gas temperature  (T$_{gas}$) &  10\,K  & 65\,K &  1, 2  \\
& Velocity dispersion ($\sigma$) & 2\,\kms & 15\,\kms &  3, 4, 5 \\
& Average volume density ($\rho$) & 10$^{2}$\,\cmc & 10$^{4}$\,\cmc & 3, 4, 5 \\
& Gas surface density ($\Sigma$) & 10$^{2}$\,\Msun\,pc$^{-2}$ & 5$\times10^{3}$\,\Msun\,pc$^{-2}$ &  6, 5  \\
\multicolumn{2}{l}{Derived:} && & \\
& Sound speed  ($c_{s}$) &  0.2\,\kms & 0.5\,\kms & \\
& Turbulent Mach number  (\mach) & 10 & 30 & \\
& Turbulent gas pressure ($P_{turb}/k$) & 10$^{5}$\,K\,\cmc  &  10$^{9}$\,K\,\cmc & \\ 
& Hydrostatic pressure from self gravity ($P_{grav}/k$)  & 10$^{5}$\,K\,\cmc  & 10$^{9}$\,K\,\cmc & \\
\hline
 & & Solar & \thebrick & \\
\hline
\multicolumn{2}{l}{Measured:} && & \\
&  {\bf {Mean, column density PDF}} (N$_{0}$) & 0.5--3.0$\times$10$^{21}$\,\cms &  86 $\pm$ 20 $\times$10$^{21}$\,\cms &  7, 8 \\
&  {\bf {Dispersion, column density PDF}} ($\sigma_{\log{N}}$) & 0.28--0.59 &  0.34 $\pm$ 0.03  &  7, 8  \\
&  {\bf {Critical volume density}} ($\rho_{crit}$) & 10$^{4}$\,\cmc & $>$ 10$^{6}$\,\cmc &  3, 9, 8 \\
\multicolumn{2}{l}{Predicted (relative to solar neighbourhood clouds): }&&  &\\
&  {\bf {Mean, column density PDF}} (N$_{0}$) & 1 & 100 & \\
&  {\bf {Dispersion, volume density PDF}} ($\sigma_{\log{\rho}}$) & 1   & 1.2 & \\
&  {\bf {Critical volume density}} ($\rho_{crit}$) & 1 & 10$^{4}$ & 10, 11, 5 \\
\hline
\end{tabular}\\
}}
\tablerefs{(1) \cite{Larson03-SFreview}; (2) \cite{Ao13}; (3) \cite{Lada10}; (4) \cite{Longmore13}; (5) \cite{Kruijssen14}; (6)  \cite{Schneider14}; (7) \cite{Kainulainen09}; (8) this work; (9) \cite{Lada12}; (10) \cite{Krumholz05}; (11) \cite{Padoan11}}\\
\end{table}

\end{document}